# Label-Free Chemical Nano-Imaging of Intracellular Drug Binding Sites.

**Authors:** William S. Hart[1,*], Hemmel Amrania[1], Alice Beckley[2], Jochen R. Brandt[3], Sandeep Sundriyal[3,4], Ainoa Rueda-Zubiaurre[3], Alexandra E. Porter[5], Eric O. Aboagye[2], Matthew J. Fuchter[3], and Chris C. Phillips[1,*].

[1]Experimental Solid State Group, Department of Physics, Imperial College London, London, SW7 2AZ, UK.
[2]Department of Surgery and Cancer, Imperial College London, London, UK.
[3]Department of Chemistry, Imperial College London, London, SW7 2AZ, UK.
[4]Department of Pharmacy, Birla Institute of Technology and Science Pilani, Pilani Campus, 333031, Rajasthan, India.
[5]Department of Bioengineering, Royal School of Mines, Imperial College London, London, SW7 2AZ, UK
*Correspondence to: chris.phillips@imperial.ac.uk

**Abstract:**
Optical microscopy has a diffraction-limited resolution of ~250nm. Fluorescence methods (e.g. PALM, STORM, STED) beat this, but they are still limited to 10's nm, and the images are an indirect pointillist representation of only part of the original object. Here we describe a way of combining a sample preparation technique taken from histopathology, with a probe-based nano-imaging technique, (s-SNOM) from the world of Solid State Physics. This allows us to image subcellular structures optically, and at a nanoscale resolution that is ~100x better than normal microscopes. By adding a tuneable laser source, we also demonstrate mid-infrared chemical nano-imaging (MICHNI) in human myeloma cells and we use it to map the binding sites of the anti-cancer drug bortezomib to <10 zL (<$10^{-20}$ L) sized intracellular components. MICHNI is *label free* and can be used with any biological material and drugs with specific functional chemistry. We believe that its combination of speed, cheapness, simplicity, safety and chemical contrast promises a transformative impact across the life sciences.

**One Sentence Summary:** Universal optical imaging of nanoscale biology and chemistry at a fraction of the effort and cost of electron microscopy.

**Introduction.**
In the past two decades a range of fluorescence cell microscopy techniques[1] have been developed which can achieve down to ~10 nm spatial resolution[2,3], *i.e.* substantially beating the usual limits (~250nm) set by optical diffraction. However, achieving this performance relies on specialised labelling[4] with bespoke "fluorescers" that have to be individually developed to suit the problem at hand. This limits the techniques' applicability as well as risking perturbing the biology[5,6]. The alternative route to sub-diffraction so-called "ultrastructural" information, electron microscopy, (EM), requires complex and time-consuming sample preparation that risks compromising the sample's integrity. Samples have to be stained with heavy metals to give usable electron contrast, and they have to be rendered into a form that is electrically conductive and can withstand the vacuum environment within the EM, a process that typically takes weeks.

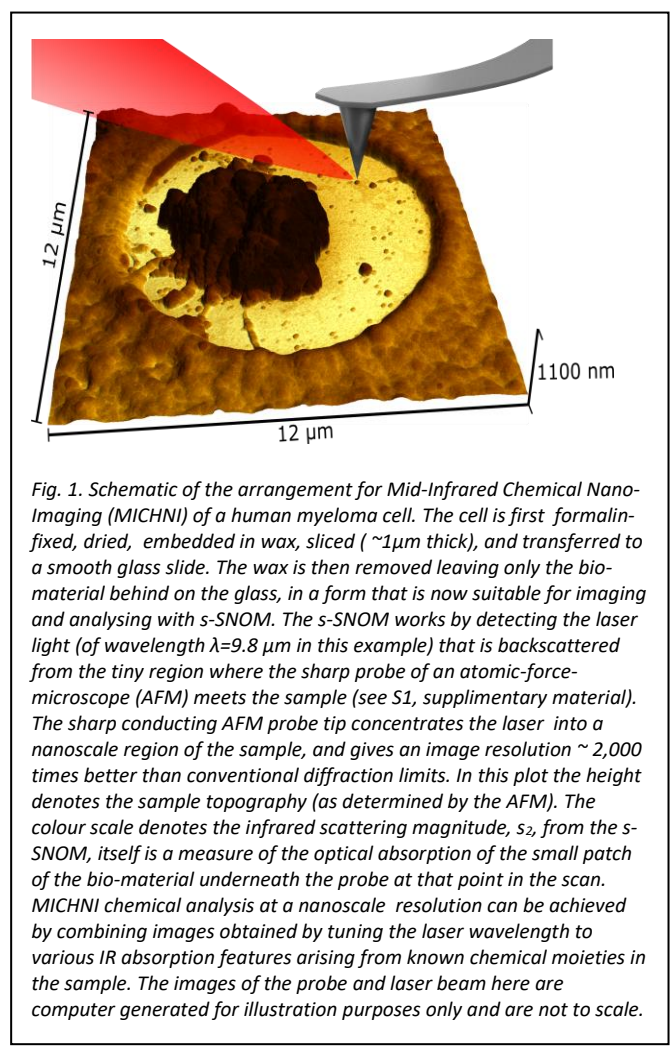

*Fig. 1. Schematic of the arrangement for Mid-Infrared Chemical Nano-Imaging (MICHNI) of a human myeloma cell. The cell is first formalin-fixed, dried, embedded in wax, sliced (~1μm thick), and transferred to a smooth glass slide. The wax is then removed leaving only the bio-material behind on the glass, in a form that is now suitable for imaging and analysing with s-SNOM. The s-SNOM works by detecting the laser light (of wavelength λ=9.8 μm in this example) that is backscattered from the tiny region where the sharp probe of an atomic-force-microscope (AFM) meets the sample (see S1, supplementary material). The sharp conducting AFM probe tip concentrates the laser into a nanoscale region of the sample, and gives an image resolution ~ 2,000 times better than conventional diffraction limits. In this plot the height denotes the sample topography (as determined by the AFM). The colour scale denotes the infrared scattering magnitude, $s_2$, from the s-SNOM, itself is a measure of the optical absorption of the small patch of the bio-material underneath the probe at that point in the scan. MICHNI chemical analysis at a nanoscale resolution can be achieved by combining images obtained by tuning the laser wavelength to various IR absorption features arising from known chemical moieties in the sample. The images of the probe and laser beam here are computer generated for illustration purposes only and are not to scale.*

In the world of physics research there are several scanning probe-based imaging methods, mostly based on atomic-force-microscopy, that have demonstrated morphological imaging resolutions[7,8] significantly below 20nm. A number of Mid-infrared (mid-IR) scanning probe techniques that are optical extensions of AFM, are also available (see S1, supplementary material). However, these only work with samples that are flat, dry and dimensionally stable down to the nm level, and they only probe down to a depth commensurate with the spatial resolution[9], so they yield essentially surface information.

These technologies have been available for more than 2 decades, and reliable, rapid and easy to use commercial instruments are now available. However, for the above reasons, thus far they have been applied only to artificial test samples, e.g. gold particles [10], or isolated proteins[11] on silicon.

Here we show how one of these probe-based techniques, scattering-type scanning near field optical microscopy (s-SNOM, Fig. 1), can be adapted to image general biological samples. The label-free imaging of cellular structures introduced in this study potentially offers a high throughput alternative to EM that is capable of aiding disease diagnosis by identifying changes in intracellular components.

As a proof-of-principle demonstration of the ability of MICHNI to locate the site of a drug's action, we here study the drug compound Bortezomib (BTZ). BTZ is an important clinically approved therapeutic against Multiple Myeloma[12], (see S5, S6 supplementary materials). It is well suited to the MICHNI approach because its chemical makeup gives it a distinctive absorption signature in the mid-IR that sets it apart from the cellular bio-material. Its mechanism of action proceeds through inhibition of proteasomes – intracellular complexes that break down proteins. As a proteasome inhibitor, one might *a priori* expect the BTZ signal to peak in regions of high proteasome concentration within the cell.

For sample preparation, we use a variant of the standard histopathological "formalin-fixed, paraffin-embedded" (FFPE) method (see S4, supplementary materials). We find that this approach can be employed to turn any biological sample into a collection of stably mounted nanoscale bio-material structures[13] that can be imaged with s-SNOM[14]. In our case, the s-SNOM (Fig. 1) is equipped with an array of widely tuneable mid-infrared quantum cascade lasers (QCLs) that span the 5 µm < λ < 11 µm range[15,16] so we can perform nanoscale spectroscopy by combining images taken over a wide range of wavelengths.

These mid-IR wavelengths lie in the so-called "chemical fingerprint" spectral region, where a given light wavelength is absorbed in a localised and bond-specific vibrational transition in the corresponding chemical moiety. These vibronic absorption features are well – characterised and have been used for many years, both for chemical analysis, and to generate chemical images using, e.g. Fourier transform infrared (FTIR) spectroscopy[17]. MICHNI exploits this link between mid-IR spectral absorption and chemical composition to allow one to infer "mid-IR chemical nano-imaging" (MICHNI) information from combinations of mid-IR s-SNOM images.

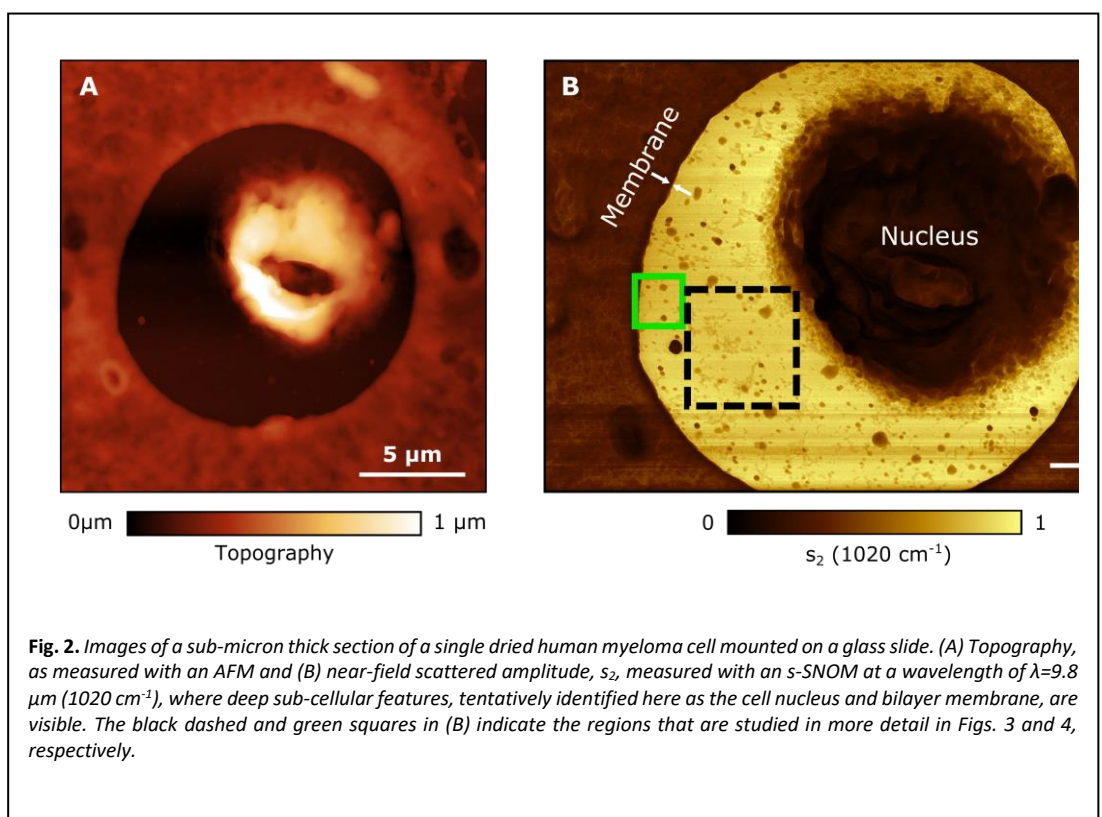

**Fig. 2.** *Images of a sub-micron thick section of a single dried human myeloma cell mounted on a glass slide. (A) Topography, as measured with an AFM and (B) near-field scattered amplitude, $s_2$, measured with an s-SNOM at a wavelength of λ=9.8 µm (1020 cm$^{-1}$), where deep sub-cellular features, tentatively identified here as the cell nucleus and bilayer membrane, are visible. The black dashed and green squares in (B) indicate the regions that are studied in more detail in Figs. 3 and 4, respectively.*

**Results.**
Figure 2 shows the AFM topography, and the second harmonic near-field scattered amplitude, $s_2$, of a section of a single RPMI-8226 myeloma cell. The cell has been pre-treated with BTZ, before being prepared for imaging (see S4, supplementary material) with the FFPE-based sample preparation protocol[13].

The AFM image (Fig. 2(A)) is essentially a height map, showing the morphological profile of the biological material in the dried section as it is distributed across the smooth glass slide. It is generated mechanically, as the probe tip scans across the sample, like a nanoscale version of a record needle. Its lateral resolution is determined by the AFM tip diameter, and there is no chemical contrast. In comparison, the MICHNI image (Fig. 2(B)) is notably sharper and resolves a host of extra sub-cellular components.

Zooming in (Fig. 3(A) and (B)) on the black dashed square of Fig. 2(B) gives a visual indication of the resolution and contrast enhancement of the MICHNI image (Fig. 3(B)) over the AFM one (Fig. 3(A)). Selecting a fine biological feature from the MICHNI image, in this case a ~15 nm high step (possibly of nuclear membrane material) at the edge of the cell nucleus (Fig. 3(D)) and performing a line scan analysis at the smallest possible step size (0.5nm), (Fig. 3(E)) shows how the way that the so-called "Lightning-rod" effect that concentrates the optical fields under the conducting tip, has resulted in a ~5x resolution improvement of the s-SNOM over AFM.

The observed ~2.3 nm MICHNI resolution corresponds to $\sim \lambda/4,300$ at the $\lambda = 9.8$ μm imaging wavelength; this beats diffraction by >3 decades. In the s-SNOM literature[9] with artificial test samples, sub ~20nm resolution figures have been reported previously. In our experiments, we found that this figure was often bettered (see supplementary material S10), but only in samples where there was strong chemical contrast, accompanied by only small height changes that were comparable to or less than the ~10 nm tip radius. The latter is estimated here from the resolutions seen in the AFM images.

If the step is higher than this, the tip has to ride up over it and, just as in AFM, this occurs over a lateral distance that is determined mechanically, by the tip radius. As the tip rides up the step, the effective tip-sample separation changes, and this purely

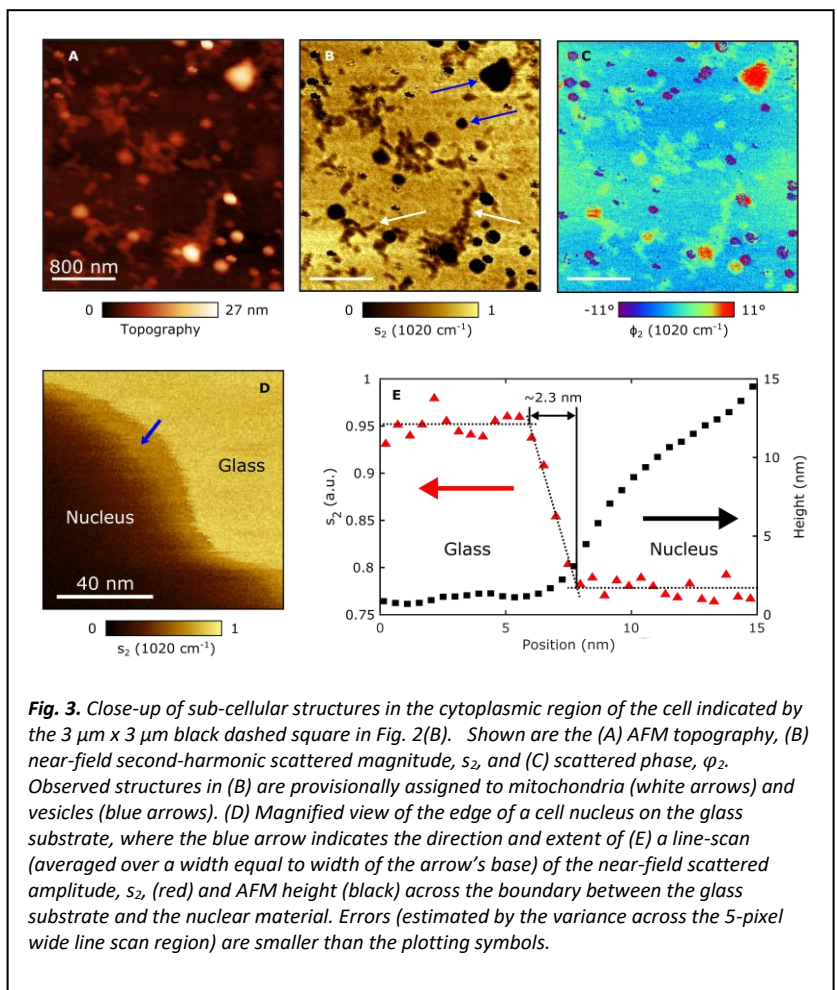

*Fig. 3.* Close-up of sub-cellular structures in the cytoplasmic region of the cell indicated by the 3 μm x 3 μm black dashed square in Fig. 2(B). Shown are the (A) AFM topography, (B) near-field second-harmonic scattered magnitude, $s_2$, and (C) scattered phase, $\varphi_2$. Observed structures in (B) are provisionally assigned to mitochondria (white arrows) and vesicles (blue arrows). (D) Magnified view of the edge of a cell nucleus on the glass substrate, where the blue arrow indicates the direction and extent of (E) a line-scan (averaged over a width equal to width of the arrow's base) of the near-field scattered amplitude, $s_2$, (red) and AFM height (black) across the boundary between the glass substrate and the nuclear material. Errors (estimated by the variance across the 5-pixel wide line scan region) are smaller than the plotting symbols.

mechanical effect alters the nano-optics that determine the backscattering cross-section.

The result is that the larger step heights will degrade the lateral resolution in the s-SNOM image to similar values to the AFM one. This mechanism may have been limiting the observed s-SNOM resolution seen in previous experiments with artificial samples because they are typically fabricated by lithography techniques, where chemical contrast is always accompanied by a height step.

The natural nanostructures studied here seem to be better for showcasing s-SNOM's true resolution limits.

The frequency-dependent phase of the backscattered light, $\varphi_2(\nu)$, is the key to the local chemical information that these images contain (see S2, supplementary material). In order to isolate the BTZ contribution to the optical response, we first take a zero-phase reference reading corresponding to a part of the image known to be glass (i.e. the lightest toned smooth parts of the $s_2$-based images). This datum is used to generate a false-colour $\varphi_2(\nu)$ image (Fig. 3(C)), where the regions of increased optical absorption

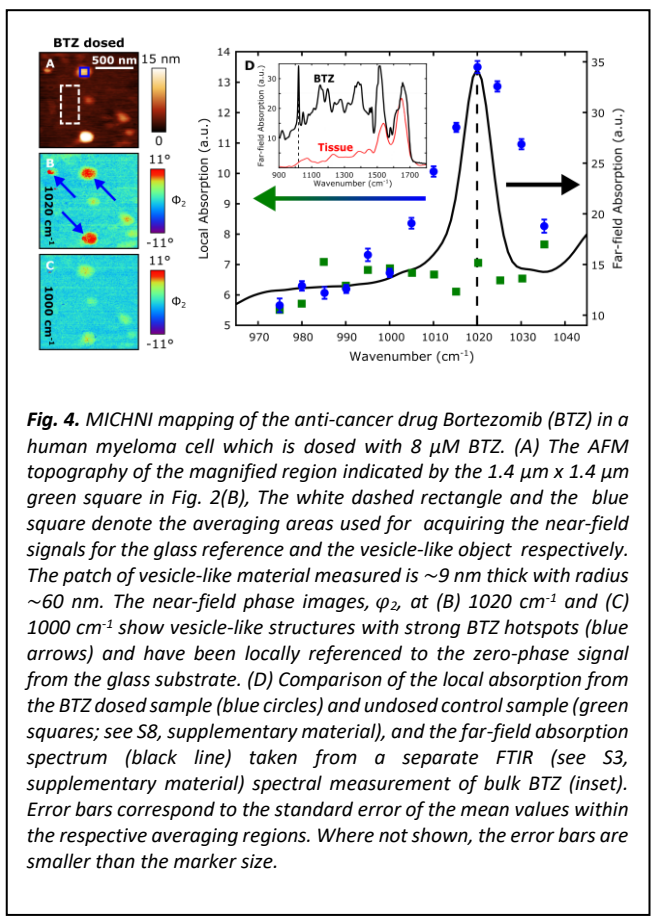

*Fig. 4. MICHNI mapping of the anti-cancer drug Bortezomib (BTZ) in a human myeloma cell which is dosed with 8 µM BTZ. (A) The AFM topography of the magnified region indicated by the 1.4 µm x 1.4 µm green square in Fig. 2(B), The white dashed rectangle and the blue square denote the averaging areas used for acquiring the near-field signals for the glass reference and the vesicle-like object respectively. The patch of vesicle-like material measured is ~9 nm thick with radius ~60 nm. The near-field phase images, $\varphi_2$, at (B) 1020 $cm^{-1}$ and (C) 1000 $cm^{-1}$ show vesicle-like structures with strong BTZ hotspots (blue arrows) and have been locally referenced to the zero-phase signal from the glass substrate. (D) Comparison of the local absorption from the BTZ dosed sample (blue circles) and undosed control sample (green squares; see S8, supplementary material), and the far-field absorption spectrum (black line) taken from a separate FTIR (see S3, supplementary material) spectral measurement of bulk BTZ (inset). Error bars correspond to the standard error of the mean values within the respective averaging regions. Where not shown, the error bars are smaller than the marker size.*

due to BTZ cause $\varphi_2(v)$ to increase and show up as red "hot spots" in the image rendering.

For quantitative spectroscopic analysis, we zoom in to an area (Fig. 4(A-C)) that corresponds to the 1.4 µm x 1.4 µm green square in Fig. 2(B). The near-field phase, $\varphi_2(v)$ image, shows strong BTZ hotspots (blue arrows) when the laser is tuned to the absorption peak of the BTZ (see S5, supplementary material). Sitting on one of these, and tuning the laser, generates an absorption peak at the same wavelength as is seen in a reference absorption spectrum (Fig. 4(D), inset) of a sample of pure BTZ taken separately in an FTIR spectrometer.

This absorption peak is due to a boron-based group in the synthetic BTZ molecule (see S5, supplementary material) that has no analogue in natural biological material. This was checked with control experiments using undosed cells (see S8, supplementary material) that generated the spectra in Fig. 4 (D) denoted by the green squares.

Although the BTZ absorption peak positions match very accurately, their widths differ because the FTIR reference spectrum is a measure of absorbed power in an unknown depth of pure drug, whereas the MICHNI signal approximates to a measure of absorption coefficient[14] in a more dilute BTZ-containing sample.

In this instance the scattering from the glass substrate is much stronger than from the biological material, so the overall scattering amplitude, $s_2$, tends to decrease with increasing thickness of the biological material[18].

The chemical contrast evident in, Fig. 4(D) is visible in features with volumes down to ~10 zL ($10^{-23}$ $m^3$).

**Discussion.**

The correspondence between the spectral signatures of the localised structures in the MICHNI images and the reference BTZ spectra argues strongly that we can map out the intracellular drug binding sites at, in this case, spatial resolutions down to ~2.3 nm.

At the same time, the images contain a wealth of nanoscale structural information arising from non-resonant mid-IR absorption in the biological material itself.

At this early stage, we are not able to identify the various intracellular elements with confidence. The various organelles are likely to shrink by unknown factors during the drying stage, and some structures may be more likely to adhere to the glass slide than others during the sample processing.

Tentatively however, we identify oval mitochondria-like structures (white arrows, Fig. 3 (B)) which, at ~250 nm long, are roughly 50% smaller than those seen in EM images of these cell lines[19]. Also present are a range of spherical vesicle-like structures with diameters in the 140 nm – 450 nm range (blue arrows, Fig. 3(B)).

Further MICHNI images of a range of different cells, which show similar reproducible cytoplasmic intracellular features, but are taken using different samples with and without BTZ, are presented in S7, supplementary material. Going forward, we believe that specialised tissue processing protocols[20] will allow the morphology and scale of these structures to be better preserved during sample preparation, and for organelle shrinkage rates to be calibrated.

Correlation experiments between MICHNI and EM would then enable these sub-cellular features to be visually identified as confidently as is done with EM at present. Similarly, correlation between MICHNI and Raman spectroscopy – a more established technique used to fingerprint spectra and map distributions of specific molecules such as lipids inside cells[21], but at much lower length scales than possible with MICHNI – could also be utilized to benchmark this technique.

In conclusion, we have demonstrated MICHNI imaging of single cells and intracellular components with a spatial resolution that already rivals EM. We speculate that future MICHNI imaging studies will prove valuable in correlating proteasome localisation and inhibition in different cell types and under different conditions, thereby informing on the sensitivity of proteasome inhibition in certain disease states (*i.e.* multiple myeloma).

A key point to note is that the FFPE sample preparation method used in MICHNI is already standard in pathology laboratories worldwide. The protocols for harvesting and preparing the tissue samples are already clinically approved and budgeted for, and reliable automated s-SNOM imagers are already commercially available. This suggests that the MICHNI technology could be readily incorporated into clinical pathology workflows in a way that could significantly improve the security and accuracy of a wide range of disease diagnoses, notably cancer.

At its core, the opto-mechanical components in the imaging system are no more complex than a DVD player, and we believe that future mass produced MICHNI equipment could be engineered to be usable and affordable in a clinical setting in a way that is not possible with EM. From the clinicians' point of view, the impact would be akin to suddenly improving their microscope resolution performance by a factor of 100. This would give them routine access to intracellular "ultrastructure" imaging information in a way that could revolutionise histopathology.

In the scientific realm, MICHNI on its own has the potential for widespread applications across the fields of cancer, cellular pathology in general, and drug discovery. It could also offer useful correlative morphological and chemical information for research problems that are currently being studied by EM.

**Materials and Methods.**

The MICHNI images presented in this study were generated using ~100 mW level tuneable (~900-1900 $cm^{-1}$) mid-infrared laser light generated by a switchable bank of commercial quantum cascade lasers (MIRcat, Daylight Solutions). The laser beam was focussed to the tip of a Pt/Ir-coated Silicon AFM probe (Arrow-NCPt, Nanoworld) in a commercial s-SNOM (NeaSNOM, Neaspec). The AFM component of the system was set to give a mean probe height of ~50 nm above the sample, and it was oscillating vertically with an amplitude of ~50 nm at its mechanical resonance frequency $\omega \approx 280$ kHz.

The sample preparation process (see S4, supplementary material) is based on the formalin-fixed-paraffin-embedded (FFPE) "gold standard" protocol[13] that is used to make many tens of thousands of biopsy sections annually in hospital pathology laboratories for routine haematoxylin and eosin ("H&E") based disease diagnoses. The process is semi-automated and takes ~ hrs (compared to ~weeks for EM sample preparation).

In most potential clinical applications, the samples would be harvested and processed this way in any event, so there is an argument that preparing clinical samples for MICHNI analysis is already clinically approved, and is effectively free and instantaneous.

The principle modifications were to use a thinner section (estimated to be ~1 μm thick before dewaxing), so that the sample height fluctuations were kept within the allowable range of the AFM, and the omission of the final H&E staining and protection stages of the sample preparation protocol altogether.

**Acknowledgments:**
Discussions with Jeremy Skepper are gratefully acknowledged.

**Funding Sources:** CCP, WH and HA acknowledge financial support from EPSRC (EP/K503733/1, EP/K029398). MF acknowledges support from EPSRC (EP/L014580/1, EP/R00188X/1) and Cancer Research UK (C33325/A19435). AR-Z acknowledges fellowship support from the Alfonso Martin Escudero Foundation. EA and AB acknowledge support from Cancer Research UK, C2536/A16584.

**Author contributions:** WH developed the MICHNI technique, acquired and analysed the MICHNI images. JB, SS and AR-Z prepared and measured far-field absorption of reference BTZ samples. AB, WH and HA supplied and prepared cell samples for MICHNI imaging. AP analysed the biology in MICHNI images. MF selected and characterised the BTZ drug. CP conceived and managed the project. All authors contributed to the preparation of the manuscript.
**Competing interests:** None of the authors have competing financial interests.

**Data and materials availability:** All data are available in the manuscript or supplementary materials.

**List of Supplementary Materials**

Additional Technical materials sections S1 – S10

Fig S1-S5

**Supplementary material:**

**S1. Technical Description and Context of Spectroscopic Mid-Infrared Chemical Nano-Imaging (MICHNI)**

Our MICHNI approach is based on so-called "scattering-type" scanning near-field optical microscopy[i] (s-SNOM). An atomic force microscope (AFM) is used to position a sharp conducting probe to within a few nm of the object to be imaged, and a laser is focussed on the region where the two meet (Fig. 1). The sharp probe tip concentrates the optical field, by a nanoscale analogue of the "lightning rod" effect, to a region commensurate with its radius.

An analysis of the backscattered light allows the optical response of the material just under the tip[i] to be measured. This gives an optical image whose spatial resolution is determined by the tip radius, and not the light wavelength[ii]. A spatial resolution of better than ~20 nm is routinely available with robust, commercially available probes. The optical image is built up by rastering the tip across the sample, and a standard AFM topography image is generated at the same time.

Previous s-SNOM studies have been restricted to narrow wavelength range studies of solid state systems. Studies include PMMA test samples[iii,iv], metamaterial superlenses[v,vi] and oxide[vii] and graphene[viii] layers. In the biological realm, demonstrator trials have imaged cylindrical tobacco mosaic viruses deposited on Si[ix], and dried proteins in lipid bilayer fragments extracted from bacteria and mounted on gold[x]. A related technique, based on AFM-photothermal effects, can yield diffraction-beating micron-scale IR images[xi,xii], but the resolution does not challenge optical microscopy and is insufficient to map out sub-cellular structure.

The MICHNI images presented in this study were generated using a NeaSNOM near field s-SNOM microscope, coupled with ~100 mW level tuneable (~900-1900 $cm^{-1}$) mid-infrared laser light generated by a switchable bank of commercial quantum cascade lasers (MIRcat, Daylight Solutions). The s-SNOM was used with a Pt/Ir-coated Silicon AFM probe (Arrow-NCPt, Nanoworld) whose tip radius is estimated as ~10nm, although the image resolution observed in this work is finer than this due to the way it concentrates the field beneath its apex[i].

The IR light backscattered from the tip was detected by a low-noise $LN_2$ cooled HgCdTe detector. The amount of scattered near-field optical signal depends on the tip-sample separation in a highly non-linear fashion, and this generates components at harmonics ($n\omega$) of the tip vibration frequency, $\omega$, in the power spectrum of the backscattered light intensity.

Analysis of the "approach curves" [ii] implies that the higher the harmonics generally contain a larger proportion of the optical near-field information, and should therefore give the best spatial resolution, with the lowest background component. However, there is a signal-to-noise trade-off because the harmonic optical power decreases with n. Signals demodulated at the $n$ = 2 harmonic were used in this study. To further decouple near-field from far-field, a pseudoheterodyne interferometric detection scheme was employed[xiii].

**S2 Extracting semi-Quantitative Spectroscopic Data from MICHNI.**

A full numerical analysis of the interaction between the laser radiation and the oscillating AFM tip is beyond the scope of this paper. However, the fact that, in solid samples, damping broadens the IR absorption peaks means that optical dispersion in the material is typically weak. This allows for an analytical model of the tip-sample interaction, (where the spherical tip is modelled as a point dipole) which, in turn, predicts that, the optical phase of the backscattered signal is proportional to the absorption coefficient of the material beneath the tip [ii,iv,ix,x,xiv].

Even if this "weakly dispersive" approximation is not reliable, given that the technique works with optical processes that are linear and therefore obey the principle of superposition, reliable concentration estimates can be obtained by measuring reference scattering phase values with the bare substrate, and subsequently recording the degree of phase shift as the tip is rastered over the bio-material [ix,x].

Small-molecule drugs are exogenous species and, in many cases, they contain chemical structures and functionality that give them a particularly distinctive mid-IR "fingerprint" spectrum. This can make them stand out from the normal absorption signature of biological material and makes them particularly suitable for mid-IR mapping (see S3).

To analyse the MICHNI images, the spectroscopic signature of the BTZ was isolated by first acquiring a frequency dependent reference phase value, $\varphi_{glass}(v)$, with the probe imaging a part of the glass substrate close to the sub-cellular structure of interest. This allowed the degree of phase shift corresponding to the BTZ absorption to be computed as $(\varphi(v) = \varphi_{BTZ}(v) - \varphi_{glass}(v))$. The backscattered amplitude was then normalised to that of the glass, $(s(v) = s_{BTZ}(v)/s_{glass}(v))$. This allowed the imaginary component of the backscattered signal, $Im(\sigma)$ to be computed as $Im(\sigma) = s \sin(\varphi)$.

$Im(\sigma)$ is itself proportional to the local absorption coefficient of the material beneath the tip[i]. For the BTZ experiments, this measure of the local absorption showed the pronounced spectral resonance at 1020 cm$^{-1}$ shown in Fig. 4(g)).

At the time of writing, this quantitative analysis cannot be performed in, for example, the depths of the cell nucleus because of difficulties in getting reliable phase calibrations due to the large height variations across the nucleus.

The quantitative spectroscopic analysis shown in Fig. 4(D) was achieved by averaging local absorption (see above) of the vesicle-like object (125 nm x 125 nm blue square in Fig. 4(A)) and normalising with respect to that of the substrate (250 nm x 500 nm white dashed rectangle). In this manner, sweeping the illumination wavelength provides a spectrum of the feature's local absorption, which can be directly compared with the known far-field absorption spectra of various chemicals (e.g. BTZ). By this means, we confirm the presence of BTZ in the vesicle-like structure imaged. The local absorption signal is averaged over a ~10 nm thick patch of material, which corresponds to a material volume of ~$10^{-23}$ m$^3$, (~10 zL).

The local absorption spectrum in Fig. 4(D) of the BTZ dosed sample (blue circles) shows good qualitative agreement with the far-field absorption spectrum (black line), which was obtained with a reference sample of pure BTZ measured in a conventional FTIR spectrometer (see S3). Both spectra show a sharp peak at 1020 cm$^{-1}$, with a difference in linewidth that is likely due to over-absorption in the reference sample whose optical thickness was not known due to the geometry of the "attenuated total reflection" method used (see S3).

**S3 Obtaining reference Bortezomib Spectra using Far-field infrared absorption spectroscopy**

A pure sample of BTZ was acquired from Selleckchem at 99.94% purity (catalogue number S1013, batch number S101315). A commercial FTIR spectrometer (Frontier FT-IR/MIR, PerkinElmer) equipped with a KBr window, an ATR (attenuated total reflectance) accessory and a LiTaO$_3$ detector was used to record the IR spectrum in the 4000-650 cm$^{-1}$ range, at 4 cm$^{-1}$ resolution and with a scan speed of 0.2 cm sec$^{-1}$.

Approximately, 2 mg of solid BTZ was applied over the exposed KBr ATR crystal surface using a carefully cleaned sample mounting plate. The knob of the pressure arm was then swung over the sample and knob was rotated to apply pressure. Finally, the ATR spectrum of BTZ was recorded (Fig. S1) and normalised against a background air spectrum.

**S4 Details of Cell preparation and Drug Treatment.**

All the studies were performed using RPMI-8226 myeloma cells, cultured in RPMI-1640 medium. The medium was supplemented with 10% FCS, penicillin (100 U/ml), 5% glutamine, and streptomycin sulphate (100 µg/ml). Cells were cultured at 37 °C in a humidified atmosphere containing 5% CO$_2$. Cells were harvested, washed twice with PBS, and counted prior to BTZ treatment. 1x10$^7$ cells were incubated with 8µM of BTZ in vehicle (0.1% DMSO) for 1 h at 37 °C. For the undosed control samples, the BTZ incubation step was omitted. Cells were washed twice in PBS before being pelleted for paraffin embedding and sectioning.

Pellets of washed RPMI-8226 myeloma cells were fixed in 4% paraformaldehyde for 30 min at room temperature. These were processed for embedding in paraffin as per published protocol[xv]. Cell blocks were cut into ~1 µm sections using a microtome before being mounted onto "charged" glass slides (Thermo Scientific product code J2800AMNZ) that had been specially treated to promote tissue adhesion. To further enhance adhesion, the glass slides were heated to 60 °C for 10 min before 5 min of cooling at room temperature. The mounted samples were dewaxed by being immersed in a series of two xylene, followed by three ethanol baths, each for 5 min.

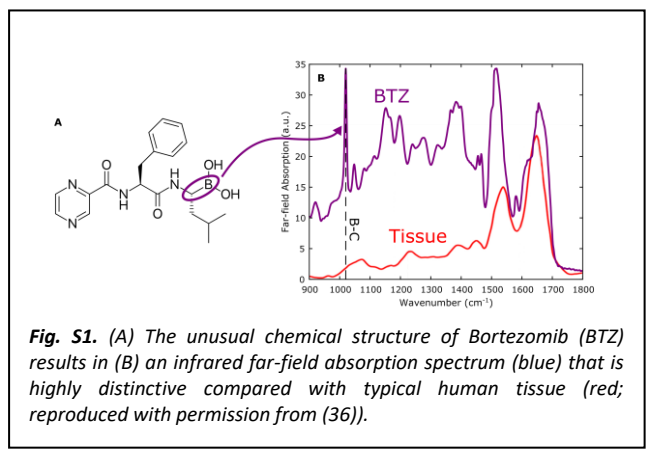

**Fig. S1.** (A) The unusual chemical structure of Bortezomib (BTZ) results in (B) an infrared far-field absorption spectrum (blue) that is highly distinctive compared with typical human tissue (red; reproduced with permission from (36)).

The MICHNI images are all taken at ambient pressure and temperature, and the cells are unstained and unlabelled throughout.

At present the specimens are imaged while they are dry, so whilst the chemical information is well preserved, the details of the sub-cellular structural morphology will be influenced by shrinkage rates that may vary between the different sub-cellular components. This makes it difficult to be confident when comparing MICHNI images with those in the EM-based literature.

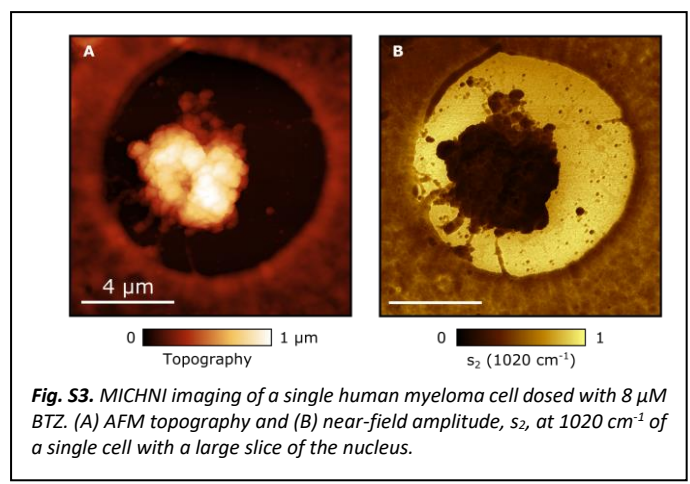

**Fig. S3.** MICHNI imaging of a single human myeloma cell dosed with 8 µM BTZ. (A) AFM topography and (B) near-field amplitude, $s_2$, at 1020 cm$^{-1}$ of a single cell with a large slice of the nucleus.

Looking ahead, we expect that a combination of operator experience, coupled with correlation studies and developments on fixing protocols will surmount this obstacle.

### S5 Comparison of Bortezomib and Human tissue mid-infrared spectra

Bortezomib (BTZ) is a promising candidate for spectroscopic mapping, as it has an unusual chemical structure, containing boron, and an associated B-C bond (Fig. S1(A)), which is rarely found in nature. This B-C bond yields a strong, sharp absorption resonance [xvi] in a spectral region where the absorption spectrum of biological material is normally featureless[xvii] (Fig. S1(B)).

### S6 Correlation between MICHNI images and the Pharmacology of Bortezomib

BTZ is a clinically approved agent for the treatment of multiple myeloma[xviii], and its mechanism of action proceeds through inhibition of proteasomes – cellular complexes that break down proteins.

As a proteasome inhibitor, one might *a priori* expect the BTZ signal to peak in regions of high proteasome concentration in the cell.

Based largely on cell-fractionation studies[xix], regions of proteasome activity and distribution are thought to typically include the outer endoplasmic reticulum, cytoskeleton and centrosomes, nuclei (but not nucleoli) and cytoplasm.

We speculate that future MICHNI imaging studies will prove valuable in correlating proteasome localisation and inhibition in different cell types and under different conditions, thereby informing on the sensitivity of proteasome inhibition in certain disease states (i.e. multiple myeloma).

### S7 Additional MICHNI images of single cells and sub-cellular structures

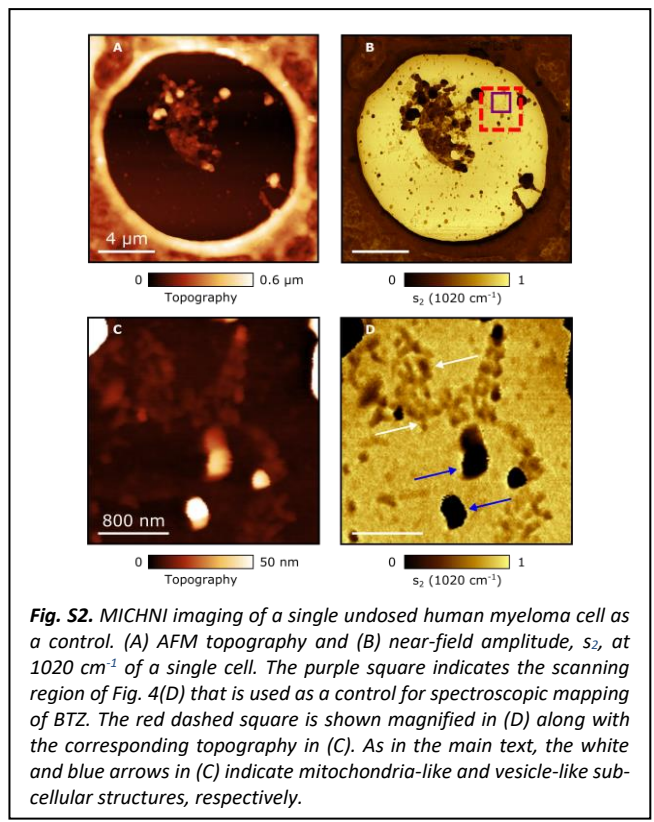

**Fig. S2.** MICHNI imaging of a single undosed human myeloma cell as a control. (A) AFM topography and (B) near-field amplitude, $s_2$, at 1020 cm$^{-1}$ of a single cell. The purple square indicates the scanning region of Fig. 4(D) that is used as a control for spectroscopic mapping of BTZ. The red dashed square is shown magnified in (D) along with the corresponding topography in (C). As in the main text, the white and blue arrows in (C) indicate mitochondria-like and vesicle-like sub-cellular structures, respectively.

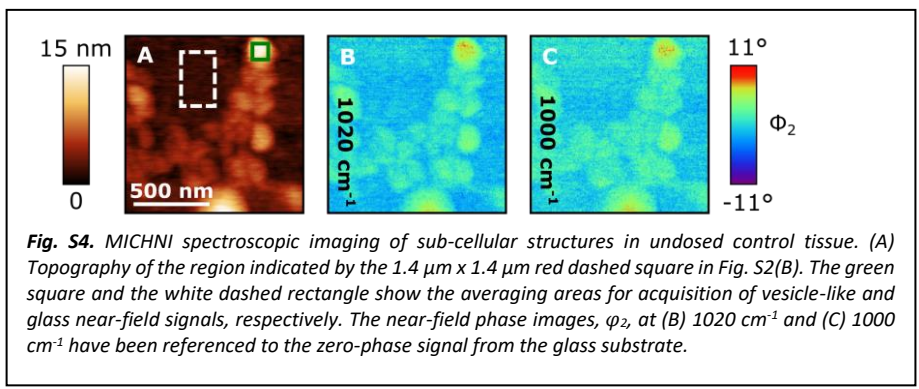

***Fig. S4.*** *MICHNI spectroscopic imaging of sub-cellular structures in undosed control tissue. (A) Topography of the region indicated by the 1.4 µm x 1.4 µm red dashed square in Fig. S2(B). The green square and the white dashed rectangle show the averaging areas for acquisition of vesicle-like and glass near-field signals, respectively. The near-field phase images, φ₂, at (B) 1020 cm⁻¹ and (C) 1000 cm⁻¹ have been referenced to the zero-phase signal from the glass substrate.*

In order to further evidence the ability of the MICHNI technique to reproducibly image the sub-cellular structure of single cells, this section provides additional images. They were taken at different times over a period of several months, in different cells, in different samples, with and without BTZ dosing. The point where the microtome has sliced the nucleus also differs significantly. In Fig. S2 only a small sliver of nuclear material has been sliced off, leaving only a few areas ≳0.6 µm high, whereas in Fig. S3 and Fig. 2 in the main text, the knife has sliced through the body of the nucleus leaving a thicker (~1 µm) residual nuclear section. However, all the images show similar cytoplasmic sub-cellular structures.

In particular, zooming in (Fig. S2(D)) reveals mitochondria-like and vesicle-like features identical to those of Fig. 3 in the main text, arguing that both are genuinely biological in origin.

### S8 Control MICHNI measurements in undosed tissue

To ensure that this spectral peak in Fig. 4(D) is not simply a signature of the endogenous chemistry of the sub-cellular structures, the spectroscopic near-field measurements were repeated with a control sample that had not been dosed with BTZ. An overview of the particular cell used in the control measurements is given in Fig. S2, which also indicates the scanning region of the control measurements that are depicted in Fig. S4(A-C). Repeating the spectroscopic analysis for this control yields the spectrum shown by the green squares in Fig. 4(D). The control does not exhibit the characteristic resonant absorption peak of BTZ, and we were unable to find any part of the control cell image that did. This is by way of confirmation that our MICHNI technique has been able to specifically map BTZ in a single nanoscale vesicle-like structure.

### S9 The relative contributions of the near-field amplitude and phase to spectral changes in Im(σ)

Strictly speaking, it is the imaginary component of the near-field scattering amplitude that is proportional to local absorption[i], however it is often the case that the majority of the chemical contrast signal in the imaginary component spectrum comes from the contribution of the near-field phase variation.

This is clearly shown in Fig. S5, whereby the amplitude spectrum is relatively flat compared with the phase spectrum. Comparing these figures with

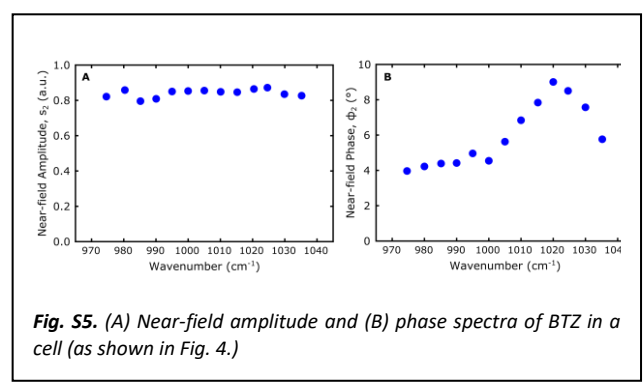

***Fig. S5.*** *(A) Near-field amplitude and (B) phase spectra of BTZ in a cell (as shown in Fig. 4.)*

Fig. 4(D) of the main text, it is clear that the phase and imaginary spectra are in good agreement.

### S10 Additional measurements evidencing sub-5 nm spatial resolution in MICHNI images

Figure S6 shows a typical linescan across the boundary between a vesicle and glass, with a ~4nm spatial resolution. In this example both the height change (~60nm), and step size (2nm) are larger than the 15nm height change/0.5nm step size in figure

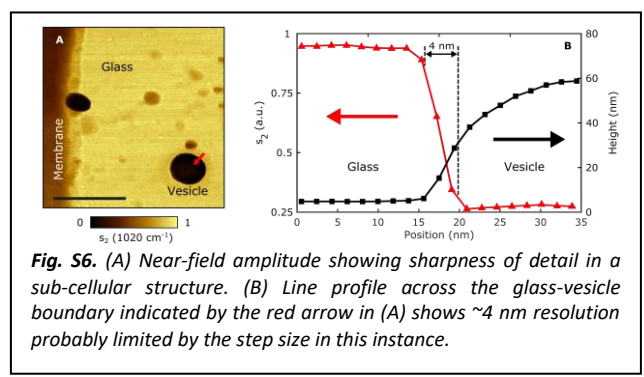

***Fig. S6.*** *(A) Near-field amplitude showing sharpness of detail in a sub-cellular structure. (B) Line profile across the glass-vesicle boundary indicated by the red arrow in (A) shows ~4 nm resolution probably limited by the step size in this instance.*

3(D) in the main text.

In many cases the s-SNOM images we observed were considerably sharper than the AFM ones. It appears that the sample area that generates the s-SNOM signal under these experimental settings, is rather smaller than the tip radius, due to the "lightning rod effect" that concentrates the light fields.

The best s-SNOM resolution figures were observed with features with strong chemical contrast (e.g. glass vs. biological tissue) and the smallest possible height changes. The especially fine detail seen in these MICHNI images argues that, as yet, nature is better at making these than man.

**Acknowledgments:**
Discussions with Jeremy Skepper are gratefully acknowledged.
**Funding Sources:** CCP, WH and HA acknowledge financial support from EPSRC (EP/K503733/1, EP/K029398). MF acknowledges support from EPSRC (EP/L014580/1, EP/R00188X/1) and Cancer Research UK (C33325/A19435). AR-Z acknowledges fellowship support from the Alfonso Martin Escudero Foundation. EA and AB acknowledge support from Cancer Research UK, C2536/A16584.
**Author contributions:** WH developed the MICHNI technique, acquired and analysed the MICHNI images. JB, SS and AR-Z prepared and measured far-field absorption of reference BTZ samples. AB, WH and HA supplied and prepared cell samples for MICHNI imaging. AP analysed the biology in MICHNI images. MF selected and characterised the BTZ drug. CP conceived and managed the project. All authors contributed to the preparation of the manuscript.
**Competing interests:** None of the authors have competing financial interests.
**Data and materials availability:** All data are available in the manuscript or supplementary materials.

**Supplementary Information References:-**